\documentclass[sigconf]{acmart}

\AtBeginDocument{%
  }

\copyrightyear{2025}
\acmYear{2025}
\setcopyright{acmlicensed}\acmConference[UniDex]{Proceedings of the xxx}{Beijing}{China}
\acmBooktitle{the Preprint Version of UniDex, Beijing, China}
\acmDOI{10.1145/xxxxx}
\acmISBN{xxxxx/2025/09}

\usepackage{amsfonts,amssymb}
\usepackage{color}
\definecolor{purplefish}{RGB}{138,43,226}
\definecolor{orangefish}{RGB}{210,105,30}
\definecolor{crimson}{RGB}{220,20,60}
\usepackage{mathrsfs}
\usepackage{url}
\usepackage{subfigure}
\usepackage{graphicx}
\usepackage{multirow}
\usepackage{makecell}

\usepackage{algorithmic}
\usepackage{algorithm}
\usepackage{amsmath} 
\usepackage{amsfonts}
\usepackage{tabularx}

\begin{document}

\title{UniDex: Rethinking Search Inverted Indexing with Unified Semantic Modeling}

\author{Zan Li$^{\dagger}$, Jiahui Chen$^{\dagger}$, Yuan Chai$^{*\dagger}$, Xiaoze Jiang$^{*\dagger}$, Xiaohua Qi$^{\dagger}$, Zhiheng Qin, Runbin Zhou, \\ Shun Zuo, Guangchao Hao, Kefeng Wang, Jingshan Lv, Yupeng Huang, Xiao Liang, Han Li}
\affiliation{%
  \institution{Kuaishou Technology, Beijing, China}
  \country{}
}

\email{{lizan07, chenjiahui11, chaiyuan, jiangxiaoze, qixiaohua03}@kuaishou.com}

\begin{abstract}
Inverted indexing has traditionally been a cornerstone of modern search systems, leveraging exact term matches to determine relevance between queries and documents. However, this term-based approach often emphasizes surface-level token overlap, limiting the system's generalization capabilities and retrieval effectiveness. To address these challenges, we propose UniDex, a novel model-based method that employs unified semantic modeling to revolutionize inverted indexing. UniDex replaces complex manual designs with a streamlined architecture, enhancing semantic generalization while reducing maintenance overhead. Our approach involves two key components: UniTouch, which maps queries and documents into semantic IDs for improved retrieval, and UniRank, which employs semantic matching to rank results effectively. Through large-scale industrial datasets and real-world online traffic assessments, we demonstrate that UniDex significantly improves retrieval capabilities, marking a paradigm shift from term-based to model-based indexing. Our deployment within Kuaishou's short-video search systems further validates UniDex's practical effectiveness, serving hundreds of millions of active users efficiently.
\end{abstract}




\maketitle

\section{Introduction}

Inverted indexing has long served as a foundational component of modern search systems due to its efficiency in large-scale information retrieval~\cite{sparterm2020, yang2017anserini, formal2021splade}. The conventional paradigm tokenizes documents into discrete terms and builds an index mapping each term to its associated documents~\cite{ formal2021splade_v2, formal2021splade}. Relevance between queries and documents is then primarily determined through exact or approximate term matches \cite{roelleke2008tf,pang2017DeepRank}. While effective in terms of efficiency, this approach inherently emphasizes surface-level lexical overlap, thereby limiting generalization and hindering retrieval effectiveness~\cite{dai2019deeper}. Breaking away from this term-centric paradigm to enhance semantic retrieval remains an open challenge.

\begin{figure}[t]
\centering
\includegraphics[width=8.5cm]{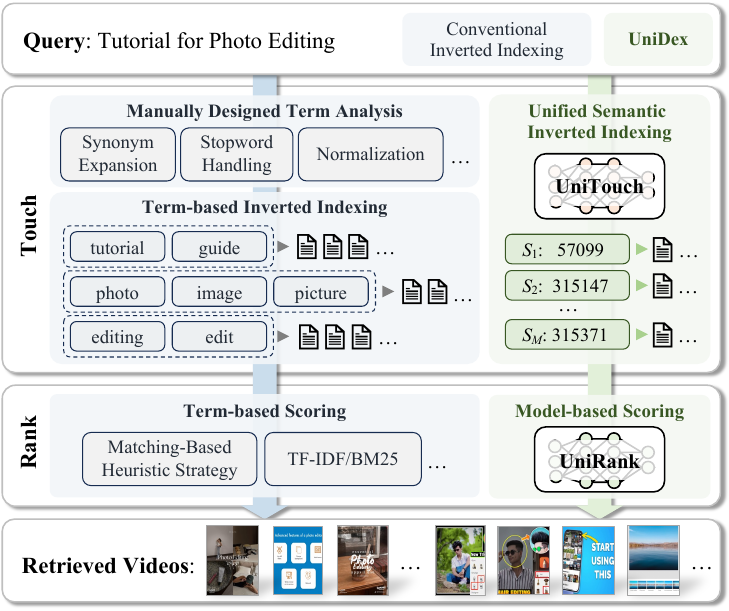}
\caption{Comparison of Conventional Inverted Indexing and UniDex. UniDex substitutes the various manually designed term-based Touch and Rank components with the unified semantic modeling approaches, UniTouch and UniRank. 
}
\label{pic:basicIdea}
\end{figure}

As illustrated in the blue blocks of Figure~\ref{pic:basicIdea}, industrial-scale inverted indexing systems are typically organized into two stages: the \textbf{Touch} stage and the \textbf{Rank} stage. The Touch module performs term-level retrieval to construct a candidate set, while the Rank module scores these candidates with respect to the query. Over the years, a variety of manually engineered heuristics have been introduced at both stages to improve recall and ranking quality~\cite{yang2017anserini,sparterm2020,bruch2024efficient}. For example, synonym expansion, stopword handling, and term normalization are employed during Touch to enlarge the candidate pool, while heuristic strategies such as BM25~\cite{yang2017anserini} or TF–IDF~\cite{roelleke2008tf}, often combined with handcrafted term-importance features, are used in Rank to refine relevance estimation~\cite{yang2017anserini, roelleke2008tf, nogueira2019document}. Despite their widespread adoption~\cite{dai2019deeper}, these heuristics are costly to maintain, brittle across domains, and fundamentally constrained by the limitations of term-level modeling. 

With the rise of deep semantic models~\cite{khattab2020colbert,jiang2022xlm,wang2025personalized}, many recent studies attempt to integrate neural representations into inverted indexing~\cite{formal2021splade,formal2021splade_v2,lassance2024splade}. However, most of these works remain tied to the term-based paradigm: semantic models are often used to generate richer sets of query or document terms, but the retrieval process itself still relies on lexical overlap. Consequently, such methods~\cite{sparterm2020, lassance2024splade} inherit the weaknesses of traditional indexing, including reliance on handcrafted matching strategies and limited semantic generalization.

Motivated by these limitations, we propose \textbf{UniDex}, a unified model-based framework that rethinks inverted indexing from a semantic perspective. As shown in the green blocks of Figure~\ref{pic:basicIdea}, UniDex replaces manually designed term-level components in both Touch and Rank with two semantic modeling modules: \textbf{UniTouch} and \textbf{UniRank}. 
Specifically, UniTouch maps queries and documents into discrete semantic IDs (SIDs) through a dual-tower encoder with an integrated quantization module. To better capture the inherent ambiguity of queries and the compositional semantics of documents, UniTouch represents each input with multiple learnable tokens, each token corresponding to a potential semantic aspect. For efficient retrieval, we design a Max–Max matching strategy that aligns query tokens with document tokens in a manner consistent with the inverted indexing lookup logic. This design ensures that a document can be retrieved if it matches at least one semantic aspect of the query, while preserving the scalability of inverted indexing. As a result, UniTouch can recall documents whose semantics are close to the query even when they do not share overlapping terms, thus significantly improving generalization over traditional term-based inverted indexes. 
On top of this retrieval stage, UniRank employs another semantic model with token-level interactions to precisely rank the retrieved candidates, going beyond conventional heuristic term-matching approaches. Together, UniTouch and UniRank form UniDex: a unified retrieval–ranking pipeline that combines the scalability of inverted indexing with the expressive power of semantic modeling. Unlike traditional pipelines, UniDex eliminates dependence on handcrafted rules, generalizes more effectively across diverse queries and domains, and substantially reduces engineering overhead, making it a practical solution for real-world search platforms.

We comprehensively evaluate UniDex on large-scale industrial datasets as well as real-world online traffic. The main contributions of this work are summarized as follows:

(1) We present the first paradigm shift of inverted indexing from a term-based to a model-based framework. This reformulation opens new directions for the evolution of retrieval systems and carries significant implications for both research and industry.

(2) We propose \textbf{UniDex}, a unified semantic modeling framework that replaces manually engineered term-matching and multi-path relevance computations with semantic ID–based indexing and model-driven ranking. This design substantially reduces maintenance overhead while providing stronger semantic generalization.

(3) To the best of our knowledge, we achieve the first successful large-scale deployment of a model-based inverted indexing system in industry. UniDex has been integrated into Kuaishou's short-video search platform, where extensive online A/B testing confirms its effectiveness and efficiency. It currently serves hundreds of millions of active users, demonstrating both the practicality and impact of our approach.

\begin{figure*}[t]
\centering
\includegraphics[width=1.0\textwidth, trim=120 150 125 150, clip]{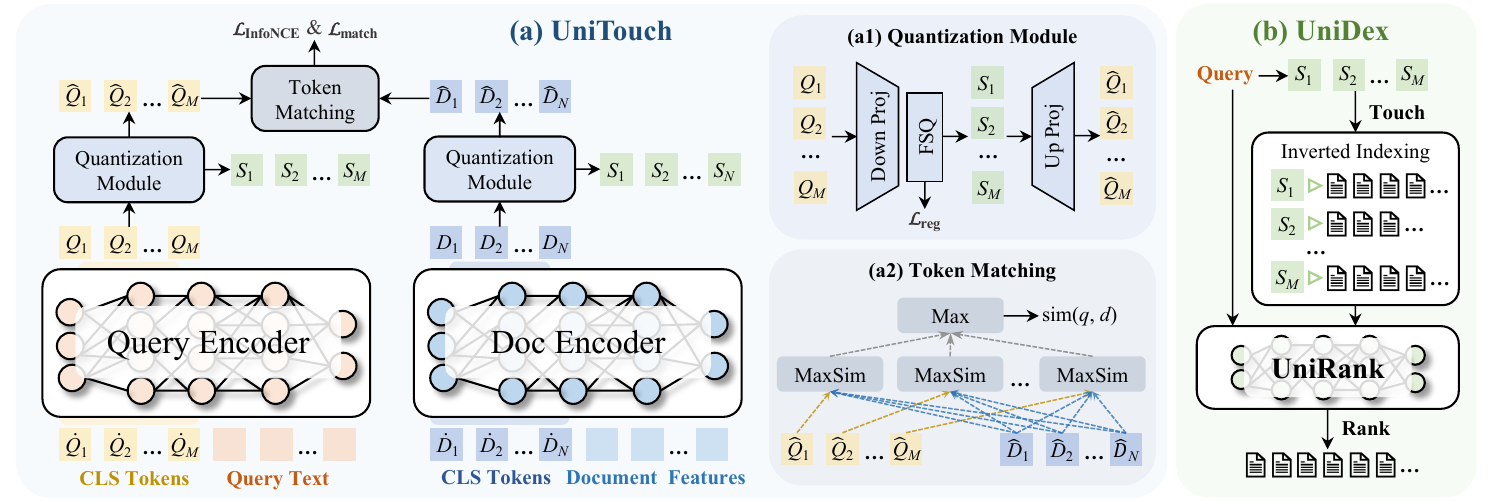}
\caption{Overview of our framework. 
(a) The UniTouch model architecture, consisting of a Query Encoder, a Document Encoder, and Quantization Modules. 
(a1) The internal design of the Quantization Module. 
(a2) The token matching strategy in UniTouch, defined as the maximum entry of the token-level matching matrix. 
(b) The end-to-end retrieval pipeline of UniDex: a query is first discretized into semantic IDs via UniTouch, relevant documents are retrieved from the semantic inverted indexing, and the final candidate set is ranked by UniRank.}

\label{fig:model}
\end{figure*}

\section{Related Works}

\subsection{Inverted Indexing}

In the field of information retrieval, inverted indexing remains the most classical and widely adopted indexing structure. Its core principle is term-based indexing, which leverages the Bag-of-Words model to establish mappings between documents and terms, thereby enabling efficient matching and retrieval. Traditional approaches such as TF-IDF \cite{roelleke2008tf} and BM25 \cite{yang2017anserini} compute query-document relevance by weighting term frequency and inverse document frequency. Despite their effectiveness, these methods inherently rely on exact lexical overlap, making them susceptible to vocabulary mismatch and unable to capture deeper semantic relationships. The rise of deep semantic modeling has driven extensive research into integrating inverted indexing with neural networks. For instance, DeepCT \cite{dai2019deeper} leverages pre-trained language models to derive context-aware term weights, enabling finer-grained term weighting.  However, it remains fundamentally limited by unresolved vocabulary mismatch. Generative approaches, such as doc2query \cite{nogueira2019doc2query} and doc2query-T5 \cite{nogueira2019document}, attempt to mitigate this limitation by predicting potential query terms to expand document representations, thus indirectly reinforcing salient terms; however, their training paradigm is inherently indirect. Another approach involves directly modeling the interactions between query or document tokens and the entire vocabulary, constructing an interaction matrix that is subsequently aggregated to obtain term-level importance scores. SparTerm \cite{sparterm2020} applies summation aggregation, while SPARTA \cite{zhao2020sparta} and EPIC \cite{MacAvaney_2020} use max pooling, yet their representations are either insufficiently sparse without additional top-$k$ pruning or lack explicit sparsity-inducing regularization. The SPLADE family of models \cite{formal2021splade, formal2021splade_v2, lassance2024splade, bruch2024efficient} remedies these limitations by combining sparsity regularization with lexical expansion, thereby improving efficiency, interpretability, and retrieval effectiveness, though the paradigm remains largely constrained by token-level interactions and heuristic term-matching rules. This reliance limits generalization and hinders the ability to fully exploit model-based representations. In contrast, UniDex, as a unified semantic modeling approach, eliminates complex handcrafted rules and enhances generalization capabilities.

\subsection{Other Retrieval Methods}
Beyond inverted indexing, dense retrieval has emerged as a complementary paradigm that encodes queries and documents into a shared embedding space for semantic matching. This typically involves an offline document encoding and indexing phase, followed by an online query encoding and similarity search stage \cite{devlin-etal-2019-bert, lee2019latent, qu2020rocketqa}. Early studies employed dual-encoder architectures built upon pretrained language models (PLMs) such as BERT \cite{devlin-etal-2019-bert, lee2019latent}, demonstrating strong performance on semantic retrieval benchmarks. Subsequent advances have sought to enhance dense retrievers through task-specific pretraining ({\it e.g.}, Condenser \cite{gao2021complement}, coCondenser \cite{gao2021unsupervised}, PAIR \cite{ren2021pair}), improved negative sampling strategies \cite{yang2024trisampler, zhan2021optimizing}, and knowledge distillation from stronger teacher models \cite{qi2025data, qu2020rocketqa}. To capture fine-grained query-document interactions, token-level and multi-representation encoders have been proposed, including ColBERT \cite{khattab2020colbert}, ColBERTv2 \cite{santhanam2021colbertv2}, ME-BERT \cite{luan2021sparse}, COIL \cite{gao2021coil}, uniCOIL \cite{lin2021few} and MVR\cite{zhang2022multi}. These methods combine expressive embeddings with efficient late interaction mechanisms or sparse representations, thereby balancing retrieval effectiveness and computational efficiency. Efficiency-oriented research has further explored embedding compression, product quantization, and binary encoding to reduce memory footprint and latency \cite{yamada-etal-2021-efficient, zhan2022learning, zhan2021jointly}. More recently, large language models (LLMs) have been leveraged for dense retrieval owing to their strong semantic understanding, multi-task adaptability, and extended context modeling, enabling more sophisticated query--document interactions and effective long-document retrieval \cite{chen2024generalizing, li2024llama2vec}. Compared with our model-based inverted indexing framework, dense retrieval depends on computationally intensive neural similarity search, while our approach eliminates manual term-matching design and benefits from the inherent efficiency of inverted search. 
In addition,  statistical retrieval methods are also used as complementary retrieval strategies~\cite{he2021purify, shi2025retrieval, gao2023alleviating}. Techniques such as item-to-item retrieval \cite{he2021purify} and collaborative filtering \cite{shi2025retrieval, fkih2023enhancing} exploit user--content associations to expand candidate sets. While effective in capturing behavioral relevance, these methods suffer from cold-start issues and the Matthew effect \cite{zhao2022improving, gao2023alleviating}, and are thus rarely deployed in isolation. Instead, they serve as auxiliary strategies when integrated into more robust retrieval frameworks. Our approach enhances inverted indexing by integrating semantic generalization capabilities, making it inherently more effective in cold-start scenarios than traditional statistical methods. This integration allows the system to better understand and relate to new and unseen queries. It elevates the potential for iterative improvements in inverted indexing.

\section{Methodology}

In this section, we first introduce the two components of UniDex: the UniTouch and UniRank models. Subsequently, we present the complete UniDex search pipeline.

\subsection{UniTouch}
\label{sec:unitouch}

\subsubsection{Model Architecture} 
The UniTouch model learns semantic representations of queries and documents, encoding them as discrete Semantic IDs (SIDs). As shown in Figure~\ref{fig:model} (a), UniTouch consists of two major components: (i) a Query/Document Encoder, which captures semantic information, and (ii) Quantization Modules, which discretizes continuous embeddings into compact symbolic identifiers.

\noindent\textbf{Query Encoder.}  
Given a user query $q$, we first tokenize it into a sequence $\{\dot{q}_1, \dot{q}_2, \ldots, \dot{q}_m\}$. To capture the polysemy of queries, we append $M$ learnable tokens $\{\dot{Q}_1, \dot{Q}_2, \ldots, \dot{Q}_M\}$ to the sequence. The encoder is instantiated as a $L$-layer BERT model. After the forward pass, we extract the embeddings of the $M$ appended tokens, yielding $\{Q_1, Q_2, \ldots, Q_M\} \in \mathbb{R}^d$, which collectively represent the semantic embedding of the query. Each token embedding captures a distinct semantic aspect of the query.

\noindent\textbf{Document Encoder.}  
The document encoder is designed analogously. For a document $d$ tokenized as $\{\dot{d}_1, \dot{d}_2, \ldots, \dot{d}_n\}$, we append $N$ learnable tokens $\{\dot{D}_1, \dot{D}_2, \ldots, \dot{D}_N\}$. The forward pass produces $N$ embeddings $\{{D}_1, {D}_2, \ldots, {D}_N\} \in \mathbb{R}^d $, with each encoding a distinct semantic component of the document. Since documents typically contain richer and more diverse semantics than queries, we set $N>M$. During training, the query and document encoders share model parameters, ensuring semantic alignment across domains.

\noindent\textbf{Quantization Module.}  
To convert continuous embeddings into discrete semantic IDs, we introduce a quantization module. The detailed network of quantization module is shown in Figure~\ref{fig:model} (a1). Taking a query embedding $Q_i \in \mathbb{R}^d$ as an example, we first project it into a lower-dimensional space:
\begin{equation}
    Q_i' = \text{DownProj} (Q_i) \in \mathbb{R}^{d_q}, \quad d_q \ll d.
    \label{eq:proj_down}
\end{equation}  
Next, we apply the \emph{Finite Scalar Quantization (FSQ)}~\cite{mentzer2023finite} algorithm, which discretizes each dimension of $Q_i'$ into one of $K$ bins:
\begin{align}
    S_i &= \text{FSQ}(Q_i') \in \{0, 1, \ldots, K-1\}^{d_q},
    \label{eq:fsq} \\
    \text{FSQ}(Q_i') &\triangleq \text{Round}[(K-1) \sigma(Q_i')],
\end{align}
where $\sigma(\cdot)$ is the sigmoid function.
Thus, each $S_i$ corresponds to one of the possible discrete codes $K^{d_q}$, which we interpret as the semantic IDs of the query.  
Finally, the discrete code $S_i$ is mapped back into the $d$-dimensional embedding space via an up-projection:
\begin{equation}
    \hat{Q}_i = \text{UpProj}(S_i) \in \mathbb{R}^d.
    \label{eq:proj_up}
\end{equation}
 This reconstructed embedding $\hat{Q}_i$ allows the model to integrate discrete IDs into downstream tasks while maintaining compatibility with continuous representation spaces.
 During optimization, we adopt the Element-wise Gradient Scaling (EWGS)~\cite{lee2021EWGS} strategy to propagate gradients through the non-differentiable quantization step. This approach mitigates training instability introduced by Straight-Through Estimator (STE)~\cite{bengio2013estimating}, leading to faster convergence and more stable learning dynamics in the quantization module.
 Empirically, we observe that setting $K = 2$ (\textit{i.e.}, binary quantization) and $d_q = 19$ yields favorable quantization performance. Under this configuration, each semantic ID corresponds to an integer within the range $[0, 2^{19}]$, balancing representation capacity and model efficiency.

\noindent\textbf{Token Matching.}  
To measure the semantic relevance between queries and documents, we adopt a ``Max-Max'' matching strategy, as illustrated in Figure~\ref{fig:model} (a2). Following the term-match essence of inverted indexing, a document should be retrieved if it matches at least one semantic aspect of the query. For instance, consider the query ``apple'', which may refer to the fruit, the technology company, or even the record label. A document discussing global technology trends may contain sections on Google, Microsoft, and Apple. Even if only the ``Apple Inc.'' aspect overlaps with the query, the document should still be retrieved. 

Formally, the similarity between the query and the document is then defined as the maximum entry of the token-level matching matrix (other matching methods detailed in Section \ref{sec:ablation}):
\begin{equation}
    \text{sim}(q, d) = \max_{i\in [M]} \max_{j \in [N]} s(\hat{Q}_i, \hat{D}_j)=\max_{i\in [M]} \max_{j \in [N]} \frac{\hat{Q}_i\cdot \hat{D}_j}{||\hat{Q}_i||\cdot||\hat{D}_j||}.
    \label{eq:sim}
\end{equation}

\subsubsection{Learning Objectives} 
To optimize the UniTouch model, we introduce multiple training objectives that jointly enhance both semantic representation learning and the alignment between SIDs of query and documents. The training data are derived from real-world search logs. Each training instance consists of a triplet $\{q, \mathcal{D}, \mathcal{Y}\}$, where $q$ denotes the user query, $\mathcal{D} = \{d_1, d_2, ..., d_n\}$ represents the set of candidate documents (including those displayed to the user as well as unexposed documents sampled as negatives), and $\mathcal{Y} = \{l_1, l_2, ..., l_n\}$ contains the associated relevance labels, which reflect the semantic relevance between queries and documents.

\noindent\textbf{Contrastive Learning.}  
We first utilize a list-wise contrastive learning strategy to refine semantic representations. For a given query $q$, we compute the InfoNCE loss over its candidate document set. Specifically, for each document $d \in \mathcal{D}$, its loss is formulated as:
\begin{equation}
    \mathcal{L}_{\text{InfoNCE}} = - \frac{1}{|\mathcal{D}|}\sum_{d_i\in\mathcal{D}}\log 
    \frac{\exp(\text{sim}(q, d_i) / \tau)}{
    \sum\limits_{d_j \in \{d_i\} \cup \mathcal{N}(d_i)} \exp(\text{sim}(q, d_j) / \tau)},
    \label{eq:infonce}
\end{equation}
where $\tau$ is a temperature parameter, and $\mathcal{N}(d_i)=\{d_k\in \mathcal{D} | l_i>l_k\}$ denotes the set of negatives. In our design, negatives include documents with lower relevance labels as well as documents from other queries within the same batch. Compared with point-wise learning, contrastive optimization captures finer-grained semantic distinctions and provides stronger supervision for representation learning.

\noindent\textbf{Matching Loss.}  
To enable UniTouch to be more effectively applied in inverted indexing structures, we further introduce a matching loss that reinforces consistency between query SIDs and the SIDs of highly relevant documents. The intuition is that if a document is highly relevant to a query, their discretized semantic identifiers should overlap in at least part of the code space. During training, this objective is only applied to documents with the highest relevance grade:
\begin{equation}
    \mathcal{L}_{\text{match}} = \frac{1}{|\mathcal{D}'|}\sum_{d_i\in\mathcal{D}'}(1-\text{sim}(q, d_i)),
    \label{eq:match}
\end{equation}
where $\mathcal{D}'=\{d_i\in\mathcal{D}| l_i=\max(\mathcal{Y})\}$ denote the set of documents with highest relevance label.

\noindent\textbf{Quantization Regularization.}  
Finally, to improve quantization stability when $K=2$, we introduce a regularization loss that encourages embeddings to stay away from the decision boundary (\textit{i.e.}, $0.5$) in each dimension:
\begin{equation}
    \mathcal{L}_{\text{reg}} = \frac{1}{M} \sum_{i\in [M]}|||\sigma(Q_i')-0.5|-0.5||_{2}^{2}.
    \label{eq:reg}
\end{equation}
This objective prevents embeddings from oscillating around the quantization thresholds, thereby stabilizing training and ensuring robust semantic discretization.

\subsubsection{Semantic Inverted Indexing}
After training the UniTouch model, we deploy it into the online search platform. The key idea is to replace traditional term-based inverted indexing with a \emph{semantic inverted indexing} constructed from the model's discrete SIDs.  

Specifically, we first perform offline inference on the document encoder to compute SIDs for all candidate items, and then build an inverted index where each entry corresponds to a particular SID. At query time, the query encoder generates SIDs for the incoming user query in real time.  
An important feature of our design is that the $M$ query SIDs represent multiple potential semantic interpretations of the query. Consequently, the retrieval process aggregates results from all $M$ query SIDs. Formally, the final candidate set is obtained as the union of documents retrieved under each query SID, ensuring broad semantic coverage and robust recall performance.

\subsection{UniRank}

Once UniTouch produces a large pool of retrieved candidates, traditional term–based ranking methods become less effective, as many of the retrieved items may not share explicit lexical overlap with the query. To address this challenge, we propose UniRank, a semantic relevance ranking model that operates on top of the semantic inverted indexing built by UniTouch. UniRank replaces multiple handcrafted lexical matching heuristics with a unified neural ranking framework, thereby simplifying the indexing pipeline and improving generalization.

\subsubsection{Model Architecture.}  
The architecture of UniRank follows a dual-tower design similar to UniTouch. Both the query and document encoders append a set of learnable tokens after their respective input sequences to capture fine-grained semantic information. Unlike UniTouch, which emphasizes broad coverage for retrieval, UniRank focuses on precise semantic alignment for ranking. Inspired by ColBERT~\cite{khattab2020colbert}, UniRank computes token-level interactions between query and document embeddings to achieve high-resolution semantic matching. Formally, given $M$ query embeddings $\{Q_1, \ldots, Q_M\}$ and $N$ document embeddings $\{D_1, \ldots, D_N\}$, the matching score is defined as:
\begin{equation}
    \text{sim}(q, d) = \sum_{i=1}^M \max_{j \in [N]} \frac{Q_i\cdot D_j}{||Q_i||\cdot||D_j||}.
    \label{eq:colbert_match}
\end{equation}
To train the UniRank model, we use a list-wise contrastive objective as described in Equation~\ref{eq:infonce} to optimize the ranking capability, and distill the fine-grained ranking scores into the model through mean squared error (MSE) loss to enhance its relevance discrimination ability.

\subsubsection{Deployment.}  
During deployment, the document encoder pre-computes embeddings for all candidate items in an offline stage, and these embeddings are stored in memory. At query time, the query encoder generates embeddings in real time. The final ranking score is computed according to Equation~\ref{eq:colbert_match}, and documents are sorted accordingly to produce the ranked results.

\subsection{UniDex}

The overall search pipeline, termed UniDex, integrates UniTouch and UniRank into a unified two-stage framework, as shown in Figure~\ref{fig:model} (b). In the first stage, UniTouch leverages semantic inverted indexing to efficiently retrieve a broad set of candidate documents. By discretizing query and document embeddings into semantic IDs, UniTouch enables scalable retrieval that captures multiple semantic interpretations of a query. This ensures high recall while maintaining efficiency comparable to traditional term-based inverted indexing.
In the second stage, UniRank takes the retrieved candidate pool and performs fine-grained semantic ranking. While UniTouch focuses on recall-oriented retrieval, UniRank emphasizes precision by computing detailed token-level interactions between queries and documents. This step filters out semantically weaker candidates and surfaces the most relevant results at the top of the list.

Compared to conventional inverted indexing, which relies on exact term matching, synonym expansion, and manually designed matching heuristics, UniDex offers a more generalizable and efficient solution. By replacing term-level indices with SIDs, UniDex eliminates the dependence on synonym rewriting and alleviates the computational overhead of handcrafted strategies. As a result, the system achieves robust generalization, scalable efficiency, and improved retrieval quality.

\begin{table*}[t]
    \caption{Performance comparison of different models on the large-scale video search dataset.}
    \label{tab:results}
    \centering
    \newcolumntype{P}[1]{>{\centering\arraybackslash}p{#1}}
    \renewcommand{\arraystretch}{1.1}
    \begin{tabular}{P{1.5cm}P{2.25cm}P{2.40cm}P{1.5cm}P{1.5cm}P{1.5cm}P{1.5cm}}
    \toprule
 &     \multicolumn{2}{c}{\textbf{Model}}&\multicolumn{2}{c}{\textbf{Recall@300(\%)}}& \multicolumn{2}{c}{\textbf{MRR@10(\%)}}\\
 \cmidrule(r){2-3} \cmidrule{4-5} \cmidrule(l){6-7}
 &     \textbf{Touch Module}&\textbf{Rank Module}&\textbf{RS}&\textbf{CS}&\textbf{RS}& \textbf{CS}\\
 \midrule

 \multirow{4}{*}[-0.2ex]{{\shortstack{\textbf{Sparse} \\ \textbf{Retrievals} }}}&  \multirow{4}{*}[-0.2ex]{\shortstack{Inverted \\ Indexing}} &{BM25~\cite{yang2017anserini}}& 49.56& 46.10& 22.21& 18.94\\
  & &{DeepCT~\cite{dai2019deeper}}& 52.05& 48.60& 23.58& 20.42\\
 & &{SPLADE~\cite{formal2021splade}}& 54.61& 50.74& 24.21& 22.91\\
 & &{SPLADE-Max~\cite{formal2021splade_v2}}& 56.56& 51.18& 25.04& 23.27\\

\midrule
\multirow{3}{*}[-0.2ex]{{\shortstack{\textbf{Kuaishou} \\ \textbf{Baseline} }}}
& {Touch-Base}&  {Rank-Base}& 55.33& 51.12& 27.50& 24.92 \\
 & {UniTouch-24L}& {Rank-Base} & 66.06& 62.45& 31.66& 26.10\\
 & {Touch-Base}&  {UniRank}& 56.24& 51.73& 29.67& 25.89 \\
   
  \midrule
  \multirow{3}{*}[-0.2ex]{{\shortstack{\textbf{UniDex} \\ \textbf{(Ours)}}}}
          & {UniTouch-6L}& &65.21&61.20&32.29&  27.13\\
          & {UniTouch-12L}& {UniRank}&68.56&63.02&33.24&28.11\\
          & {UniTouch-24L}& &\textbf{70.74} & \textbf{65.80}&\textbf{34.06}& \textbf{28.42}\\
\midrule
\midrule
  \multirow{4}{*}[-0.2ex]{{\shortstack{\textbf{Dense} \\ \textbf{Retrievals} \\ \textbf{(For Refer)}} }}
	&\multicolumn{2}{c}{DPR~\cite{karpukhin2020dense}}& 69.57& 64.38& 34.08& 28.11\\
    &\multicolumn{2}{c}{ANCE~\cite{xiong2020approximate}}& 70.02& 65.73& 34.56& 28.62\\
   &\multicolumn{2}{c}{ColBERT~\cite{khattab2020colbert}}& 70.98& 66.16& 34.72& 29.10\\
   &\multicolumn{2}{c}{TriSampler~\cite{yang2024trisampler}}& \textbf{73.09} & \textbf{67.75}& \textbf{35.27} & \textbf{29.96} \\

 \bottomrule
 \end{tabular}
\end{table*}

\section{Experiments}
In this section, we present the implementation details, and both offline and online experimental analyses related to UniDex.

\subsection{Implementation Details}
\noindent \textbf{Datasets.}  
To rigorously evaluate UniDex, we construct large-scale training and testing datasets derived from real-world search logs of the Kuaishou App, ensuring diversity and representativeness across user behaviors. Specifically, the training dataset comprises approximately 120 million user search sessions, which cover over 3 billion query-video pairs and span multiple stages of the online search pipeline, including recall, pre-ranking, and ranking. Each session contains up to 30 candidate videos, which are uniformly sampled across different stages and divided into multiple tiers to better capture varying degrees of relevance. Hard negative samples are sampled from the earlier stages of the search pipeline, while high-quality positive samples are constructed by integrating scores from the online fine-grained ranking model with explicit user feedback. Each query-video session is represented with rich features, including textual features of both queries and videos, video consumption data, and user feedback signals. 
For offline evaluation, we select 10 million videos from the video library to form the large-scale candidate pool and extract 50,000 user search sessions from the online system to construct the test set. This design allows us to rigorously assess retrieval and ranking performance under realistic and scalable conditions.

\noindent \textbf{Details of the UniDex.}
The UniTouch encoder is implemented using a 24-layer BERT~\cite{devlin-etal-2019-bert}, initialized with internally pre-trained weights to leverage domain-specific knowledge. The model employs a hidden dimension of 1024, and training is conducted with a batch size of 32 using the Adam optimizer. We adopt an initial learning rate of $2 \times 10^{-5}$, with a linear warm-up for the first 2,000 steps, followed by a cosine decay schedule. Maximum sequence lengths are set to 32 for queries and 256 for videos, balancing computational efficiency with sufficient contextual coverage. 
Semantic information is encoded through 19-dimensional 2-bit quantized vectors, with 3 SIDs generated per query and 8 SIDs per video, enabling fine-grained semantic alignment between queries and video candidates. For the UniRank module, the settings of all training-related experimental hyperparameters are consistent with those adopted in UniTouch, and UniRank encodes the semantic information of both the query and the video into four 128-dimensional dense vectors each.

\noindent \textbf{Evaluation Metrics.}
Following \citeauthor{chen2025unisearch} \cite{chen2025unisearch}, to quantitatively assess retrieval and ranking performance, we adopt Recall@300 and Mean Reciprocal Rank (MRR@10) as our primary evaluation metrics. Recall@300 evaluates the ability of the model to retrieve relevant candidates within the top 300 results, while MRR@10 measures the ranking quality by considering the position of the first relevant item, truncated at the top 10 ranks. 
Formally, for Recall@300, let $t_i$ denote the number of true positive instances among the top 300 retrieved results for the $i$-th query, $y_i$ denote the total number of positive instances associated with that query, and $n$ denote the total number of queries. Recall@300 can be written as: $\textnormal{Recall@300} = \frac{1}{n} \sum_{i=1}^{n} \frac{t_i}{y_i}$. For MRR@10, let $t_i$ denote the rank position of the first relevant item within the top 10 retrieved results for the $i$-th query (with $t_i = \infty$ if no relevant item appears in the top 10). MRR@10 can be denoted as:
$\textnormal{MRR@10} = \frac{1}{n} \sum_{i=1}^{n} \frac{1}{t_i}$. 
We conduct evaluations on two distinct subsets: the \textit{ranking subset} (RS) and the \textit{click subset} (CS). The RS test set comprises videos recommended to users by the search system and is used to evaluate alignment with system-level preferences. In contrast, the CS test set treats user-clicked items as positives, capturing immediate user interest and reflecting real-world interaction signals. This dual evaluation strategy enables a comprehensive analysis of both systemic relevance and user-centric engagement.

\subsection{Main Results}
To assess the effectiveness of UniDex, we carry out comprehensive experiments on the large-scale video search dataset and compare its performance with existing sparse retrieval methods, dense retrieval models, and the inverted indexing baseline currently used by Kuaishou production (Online Benchmark). All experiments are conducted under consistent training and testing conditions to ensure a fair comparison.

\noindent \textbf{Compared Methods.}  
We consider the following three categories of models:  
(1) {\it Sparse Retrieval.}  
The compared sparse retrieval models include BM25~\cite{yang2017anserini} and its neural extensions. DeepCT~\cite{dai2019deeper} improves lexical matching by estimating the contextual importance of individual terms. More recent advances, including SPLADE~\cite{formal2021splade} and SPLADE-Max~\cite{formal2021splade_v2}, leverage sparsity-inducing regularization combined with lexical expansion to achieve a competitive trade-off between retrieval accuracy and efficiency.  
(2) {\it Online Benchmark.}  
To evaluate the experimental results against the baseline in a real production environment, we compare UniDex with the inverted indexing retrieval framework currently deployed in Kuaishou's search system. This framework consists of two modules, referred to as {\it Touch-Base} and {\it Rank-Base} in our experiments.
(3) {\it Dense Retrieval.}  
Finally, we introduce existing dense retrieval methods. It is important to note that dense retrieval fundamentally differs from inverted indexing, yet both serve as crucial recall mechanisms in search systems. We provide a comparison of complex dense retrieval methods as a reference for evaluation. 
DPR~\cite{karpukhin2020dense} uses a dual-encoder framework to map queries and passages into low-dimensional dense vectors, enabling efficient retrieval via dot-product similarity. ANCE~\cite{xiong2020approximate} introduces approximate nearest neighbor contrastive learning to address the training bottleneck in dense retrieval. ColBERT~\cite{khattab2020colbert} enhances query and document representations with a token-level late-interaction mechanism, enabling multi-vector retrieval. TriSampler~\cite{yang2024trisampler} further refines hard negative sampling strategies.

\noindent \textbf{Advantages of Unified
Semantic Modeling.}  
On the one hand, UniDex achieves significant growth compared to sparse models. As demonstrated in the {\it Sparse Retrievals} block of Table ~\ref{tab:results}, even our lightweight model (UniTouch-6L and UniRank) outperforms the strongest sparse methods, achieving an 8.65\% improvement in Recall@300 and a 7.25\% improvement in MRR@10 on the RS dataset. On the other hand, at its peak performance, UniDex can match the effectiveness of dense retrieval methods. As shown in the {\it Dense Retrievals} block of Table~\ref{tab:results}, our top-performing UniDex model (UniTouch-24L and UniRank) lags behind the leading model by 2.35\% in Recall@300 and 1.21\% in MRR@10 on the RS dataset. This outcome is consistent with our expectations. Dense retrieval and inverted indexing recall represent two distinct retrieval strategies. Dense retrieval is more complex and demands greater computational resources. Additionally, it is more sensitive to increases in retrieval scale, such as the expansion of candidate sets, compared to inverted indexing recall. The above analysis demonstrates that through unified semantic modeling, UniDex significantly enhances the recall capability of inverted indexing retrieval, offering a new approach for this retrieval method.

\noindent \textbf{Effectiveness of the Comparison with Online Benchmark.} 
We further compare UniDex with Kuaishou's strongest online product baseline, displayed in the {\it Kuaishou Baseline} block of Table \ref{tab:results}. Let's take Recall@300 on RS as an example for analysis. Compared to {\it Touch-Base} and {\it Rank-Base}, our comprehensive model shows a 15.41\% improvement. This significant enhancement in metrics is due to our unified semantic modeling, which improves the semantic understanding capabilities of inverted indexing retrieval and provides a robust alternative to traditional inverted indexing methods. For a more detailed analysis, applying UniTouch alone without modifying the rank module results in a 10.73\% increase. However, there's still a 4.68\% gap compared to UniDex. At the same time, replacing only the {\it Rank-Base} with UniRank yields a gain of 0.91\%. This suggests that the UniTouch module is more critical, as it can access more semantically relevant results for the rank module. Furthermore, the unified semantic modeling of UniTouch and UniRank can collaborate to achieve optimal results.

\noindent \textbf{Expanding the Iteration Space of Inverted Indexing.} As shown in the {\it UniDex} block of Table \ref{tab:results}, we explore various configurations of the UniTouch module (maintain UniRank at a consistent 24 layers) and observe that as the number of model layers increases, the metrics consistently show improvement (we set UniTouch-24L and UniRank as our online experiment setting). This indicates that further iterations of UniDex could yield even greater business benefits. It creates opportunities for continuous iterations of inverted indexing, thereby raising the business potential. Meanwhile, the unified modeling approach of UniDex removes the necessity for extensive manual term expansion and the design of manual term matching relevance methods. This streamlines the iteration process and significantly reduces maintenance costs.

\begin{table}[t]
\centering
\renewcommand{\arraystretch}{1.1}
\caption{Comparison of various token-match mechanisms.}
\begin{tabular}{l*{6}{c}}
\toprule 
\multirow{2}{*}[-0.75ex]{\textbf{Method}} & \multicolumn{2}{c}{\textbf{Recall@300}(\%) } & \multicolumn{2}{c}{\textbf{MRR@10}(\%) }\\
\cmidrule(r){2-3} \cmidrule(l){4-5}
 & \textbf{RS} & \textbf{CS} & \textbf{RS} & \textbf{CS} \\
\midrule 
UniDex-Max-Sum & {34.58} & {32.19} & {20.10} & {17.67} \\
UniDex-Max-Mean & {42.41} & {39.60} & {23.73} & {19.83} \\
UniDex-Max-Max & \textbf{70.74} & \textbf{65.80} & \textbf{34.06} & \textbf{28.42} \\
\bottomrule 
\end{tabular}
\label{tab:tt_ablation}
\end{table}

\begin{table}[t]
\centering
\renewcommand{\arraystretch}{1.1}
\caption{Results of the influence of learning objectives, where ML represents the Matching Loss, and QR represents the Quantization Regularization.}
\begin{tabular}{l*{6}{c}}
\toprule 
\multirow{2}{*}[-0.75ex]{\textbf{Method}} & \multicolumn{2}{c}{\textbf{Recall@300}(\%) } & \multicolumn{2}{c}{\textbf{MRR@10}(\%) }\\
\cmidrule(r){2-3} \cmidrule(l){4-5}
 & \textbf{RS} & \textbf{CS} & \textbf{RS} & \textbf{CS} \\
\midrule 
UniDex  & \textbf{70.74} & \textbf{65.80} & \textbf{34.06} & \textbf{28.42} \\
\cmidrule(r){1-5}
UniDex w/o ML & {67.88} & {63.05} & {33.59} & {27.96} \\
UniDex w/o QR & {70.62} & {65.54} & {34.01} & {28.25} \\

\bottomrule 
\end{tabular}
\label{tab:loss_ablation}
\end{table}

\subsection{Ablation Study}
\label{sec:ablation}

\begin{table}[t]
\centering
\renewcommand{\arraystretch}{1.1}
\caption{Analysis results of the FSQ codebook space on the experimental effects, where FSQ vectors are 2-bit quantized vectors with dimension $d_q$, and the number of SIDs on the query and video sides are 3 and 8, respectively. ``*'' indicates the setting we adopted.}
\begin{tabular}{l*{7}{c}}
\toprule 
\multirow{2}{*}[-0.5ex]{\textbf{Method}} & \multirow{2}{*}[-0.5ex]{\parbox[c]{0.5cm}{\centering \textbf{$d_q$}}} & \multicolumn{2}{c}{\textbf{Recall@300}(\%) } & \multicolumn{2}{c}{\textbf{MRR@10}(\%) } & \multirow{2}{*}[-0.5ex]{\parbox[c]{1.0cm}{\centering \textbf{Recall\\Num.}}}\\
\cmidrule(r){3-4} \cmidrule(l){5-6}
 & & \textbf{RS} & \textbf{CS} & \textbf{RS} & \textbf{CS} \\
\midrule 
 \multirow{6}{*}[-0.2ex]{{\shortstack{UniDex}}} & 16 & \textbf{71.60} & \textbf{66.23} & \textbf{34.18} & \textbf{28.47} & {26K} \\
 & 18 & {70.84} & {65.91} & {34.03} & {28.39} & {19K} \\
  & 19{*} & {70.74} & {65.80} & {34.06} & {28.42} & {16K} \\
& 20 & {69.81} & {64.78} & {33.75} & {28.08} & {13.5K} \\
& 22 & {68.76} & {63.38} & {33.48} & {27.71} & {10K} \\
& 24 & {66.16} & {61.92} & {32.58} & {27.11} & {7.6K} \\

\bottomrule 
\end{tabular}
\label{tab:ablation_fsq_codebook}
\end{table}

We perform ablation studies to assess the impact of several key configurations in UniDex on final performance, with particular emphasis on the token-matching mechanism, training strategy, dimensionality of the FSQ codebook space, and the number of SIDs produced by the query and video.

\noindent \textbf{Effect of Token-Matching Mechanisms.}
The token-matching mechanism of UniDex is crucial for adapting to inverted indexing, as detailed in Section \ref{sec:unitouch}. We evaluate the impact of various token-matching objectives between query-video pairs on performance, including Max-Sum ($sim(q,d) = \sum_i^{M}\max_{j\in [N]}s(\hat{Q}_i, \hat{D}_j)$), Max-Mean ($sim(q,d) = \frac{1}{M}\sum_i^{M}\max_{j\in [N]}s(\hat{Q}_i, \hat{D}_j)$), and Max-Max ($sim(q,d) = \max_{i\in [M]}\max_{j\in [N]}s(\hat{Q}_i, \hat{D}_j)$). The results in Table~\ref{tab:tt_ablation} evident that the Max-Max interaction significantly outperforms the other methods. This is mainly because Max-Max better preserves consistency between the training and retrieval phases of the UniTouch, and aligns well with the inverted indexing paradigm. In inverted indexing, retrieval occurs only when at least one SID from either the query side or the video side matches. In contrast, Max-Sum and Max-Mean require that all SIDs on the query side reach a certain level of similarity with SIDs on the video side to enable retrieval, which deviates from the mechanism of inverted indexing.

\noindent \textbf{Influence of Learning Objectives.} 
Given that contrastive loss is the main optimization objective in the retrieval domain and has been extensively explored in prior research \cite{oord2018representation,magnani2022semantic}, our focus here is on examining the effects of {\it matching loss} and {\it quantization regularization}. Matching loss (ML) ensures that similar queries and videos receive similar SIDs. When disabled ML, shown in Table~\ref{tab:loss_ablation}, it (UniDex w/o ML) leads to a 2.86\% decrease in Recall@300 on the RS. The sequence matching of SIDs for semantically similar content significantly affects performance. This is especially crucial in inverted indexing, where SIDs should contain as many identical values as possible for similar semantics. Then, removing quantization regularization (QR) from UniDex (UniDex w/o QR) results in a reduction of a fraction of a percent across the metrics. Although the reduction in metrics is minor, it holds significant importance in engineering practice. Variations in model training precision and the numerical precision of online inference optimizations can lead to uncontrollable changes in the decimal places of float values in the results. During quantization, errors may be incorrectly mapped to 0 or 1 (since we utilize binary quantization, detailed in Section \ref{sec:unitouch}), negatively impacting online applications. Thus, we design this regularization to keep the model away from the decision boundary of 0.5, mitigating uncontrollable changes caused by device precision issues.

\begin{figure}[t]
\centering
\includegraphics[width=8.2cm]{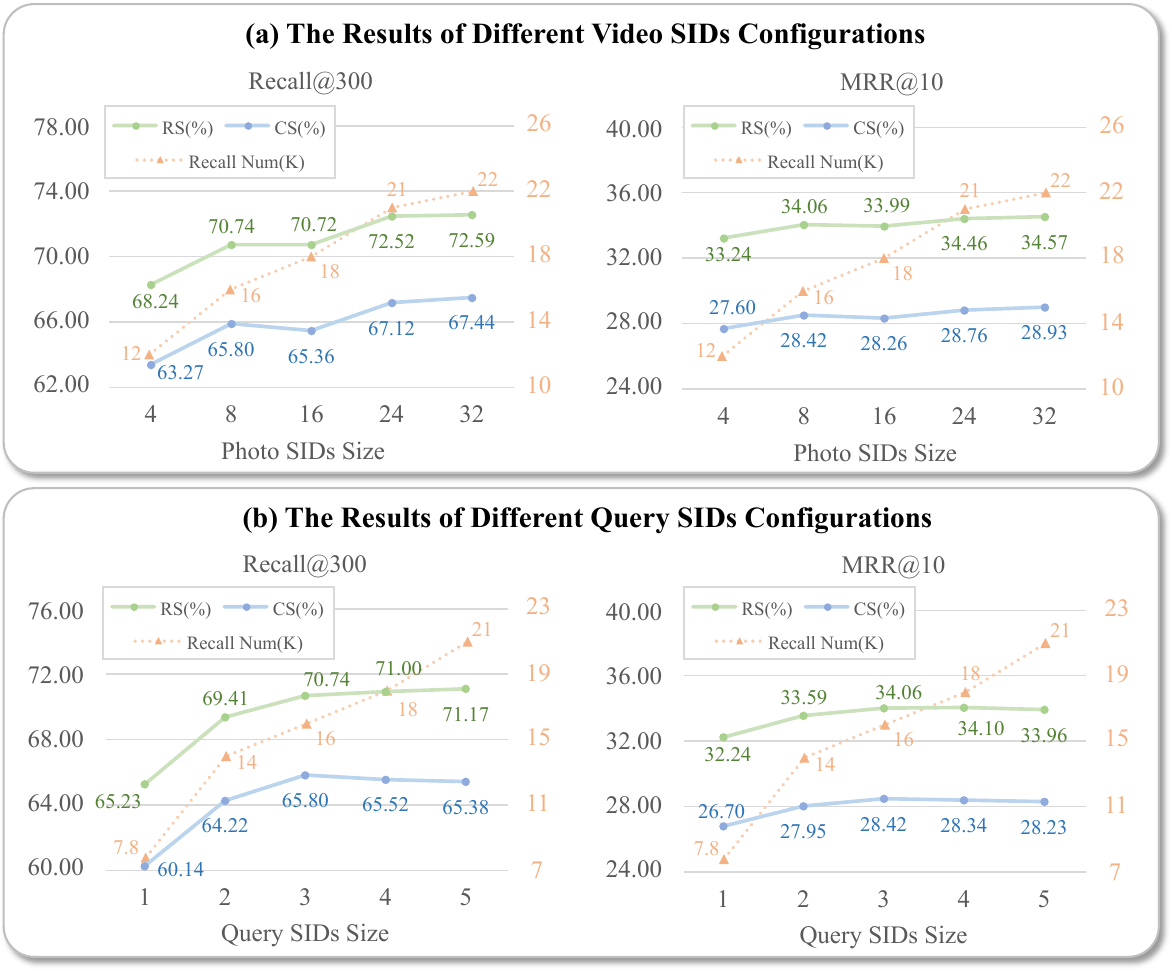}
\caption{The results of configurations with different SIDs quantities on RS and CS test set. (a) Comparison results of different SIDs quantities for Video, and (b) Comparison results of different SIDs quantities for Query.
}
\label{pic:sid_count_analyse}
\end{figure}

\noindent \textbf{Analysis on FSQ Codebook Space.}  
The FSQ's quantized codebook space significantly affects both the effectiveness of semantic models and the quality of inverted indexing. To clearly explain these effects, we conduct a series of controlled experiments, as detailed below. To ensure that our experiments are comparable, we keep the number of SIDs constant for both queries and videos in all tests. Specifically, we create six experimental groups with FSQ codebook spaces ranging from $2^{16}$ to $2^{24}$ ({\it i.e.}, $d_q$ from 16 to 24). 
As reflected in Table~\ref{tab:ablation_fsq_codebook}, there is a clear trend: the Recall@300 metric gradually decreases as the codebook space grows. This is primarily due to the rapid expansion of the codebook's quantization dimensions, which causes the semantic space to become more dispersed. Such dispersion, in turn, reduces the recall rate during the retrieval phase to some extent. Conversely, enhancing recall is associated with an increase in the number of retrieval results. For instance, when $d_q=16$, the recall results peak at 26K. This growth in recall results significantly adds to the system's overall  burden, leading to higher latency and increased consumption of computational resources. To achieve a better balance between retrieval rate and system efficiency, we set $d_q=19$. At this dimension, the decrease in Recall@30 and MRR@10 compared to $d_q=16$ is minimal, while the number of retrieved items is significantly reduced by 10K.

\noindent \textbf{Impact of SIDs Count on Query/Video.}  In the UniDex framework, the number of SIDs associated with queries and videos significantly impacts both model performance and online retrieval efficiency. On the query side, the number of SIDs directly impacts the scope of recall. Since queries usually consist of short, semantically clear text, we empirically investigate configurations with 1 to 5 SIDs. On the contrary, on the video side, the text is generally longer and contains rich semantic information, prompting us to explore settings ranging from 4 to 32. As shown in Figure \ref{pic:sid_count_analyse} (a), we start by examining the impact of Video SIDs size on the results. Clearly, as the size of SIDs increases, both Recall@300 and MRR@10 improve. Additionally, it can be observed that increasing the size from 8 to 24 (a threefold increase) results in an approximate 2\% improvement in Recall@300 and about a 0.4\% improvement in MRR. However, this change also results in an increase of 5K in the number of retrieved items, which can significantly raise inference resource consumption in the ranking module. Therefore, we select 8 as our final setting. Subsequently, Figure \ref{pic:sid_count_analyse} (b) demonstrates how the query-side SIDs size affects the outcomes. We notice that when the size surpasses 3, further improvements in metrics are minimal. This is likely because queries are typically short, and their semantic information can be effectively captured by just 3 SIDs.

\begin{table}[t]
\centering
\renewcommand{\arraystretch}{1.1}
\caption{The results of UniDex in online A/B test compared to the production baseline. We consider the user's satisfaction (Sat.) and resource costs (Cost).}
\begin{tabular}{l*{5}{c}}
\toprule 
 \multirow{5}{*}[-0.5ex]{\textbf{UniDex}} & \multirow{2}{*}[-0.5ex]{\textbf{Sat.}} & \textbf{CTR} $\uparrow$ & \textbf{VPD} $\uparrow$ & \textbf{LPC} $\uparrow$ & \textbf{MRS} $\uparrow$ \\
 \cmidrule(r){3-6} 
 &  & +0.185\%  &  +0.287\%  &  +0.352\% & +0.346\%   \\
\cmidrule(r){2-6} 
  & \multirow{2}{*}[-0.5ex]{\textbf{Cost}} &  \multicolumn{4}{c}{\begin{tabularx}{0.5\linewidth}{ccc}
\textbf{Core} $\downarrow$  & \textbf{Memory} $\downarrow$  & \textbf{Latency} $\downarrow$  \\
\end{tabularx}} 
 \\
\cmidrule(r){3-6} 
   & &  \multicolumn{4}{c}{\begin{tabularx}{0.5\linewidth}{ccc}
  -20550 \ \  &  \ \  -37TB   \ \  &   \ \ \ \  \ \ \   -25\%  \ \ \\
\end{tabularx}} 
 \\
\bottomrule 
\end{tabular}
\label{tab:oneline_res}
\end{table}

\subsection{Online Testing}
\label{sec:online_test}

We further evaluate UniDex in real-world settings by deploying it within Kuaishou's short-video search system. A 5-day online A/B test was conducted, with both the experimental and control groups randomly assigned 10\% of the actual search traffic.
  
\noindent \textbf{Improving User Satisfaction with Search Results.} We focus on four online metrics: page click-through rates (CTR), video playback duration (VPD), long play count (LPC), and the mean relevance score of the top 4 results (MRS). As shown in {\it Sat.} block of Table \ref{tab:oneline_res}, UniDex outperforms the advanced production baseline in terms of CTR, VPD, and LPC. This indicates that users are clicking on more videos and spending more time watching relevant content, significantly improving their experience. Additionally, MRS serves as a key monitoring metric, reflecting changes in the fine-grained relevance scores of the top four results shown to users after implementing our model. A MRS improvement of +0.346\% suggests that UniDex can deliver more relevant videos compared to traditional term-based inverted indexing systems. This enhancement not only improves the system's overall relevance representation capability but also positively influences the search system's long-term evolution.

\noindent \textbf{Minimizing System Resource Expenditure.}  UniDex has replaced the traditional term-based inverted indexing used online. In the conventional pipeline, manual term matching and relevance computation consume significant resources. With the full deployment of UniDex, a substantial amount of inference resources has been saved. As shown in the {\it Cost} block of Table~\ref{tab:oneline_res}, it saves 20550 cores and 37TB of storage resources, resulting in a 25\% improvement in system response time. The reduction in computational and storage demands leads to lower operational costs, allowing for more efficient allocation of resources. This efficiency enables the system to handle higher volumes of concurrent user requests, effectively increasing its scalability and robustness. Furthermore, the 25\% improvement in system response time enhances the user experience by providing faster access to relevant information, which can boost user engagement and satisfaction.

\subsection{Further Analysis}
\label{sec:analysis}

To better understand the advantages of UniDex over traditional term-based inverted indexing, we conduct a more in-depth analysis.

\begin{figure}[t]
\centering
\includegraphics[width=8.7cm]{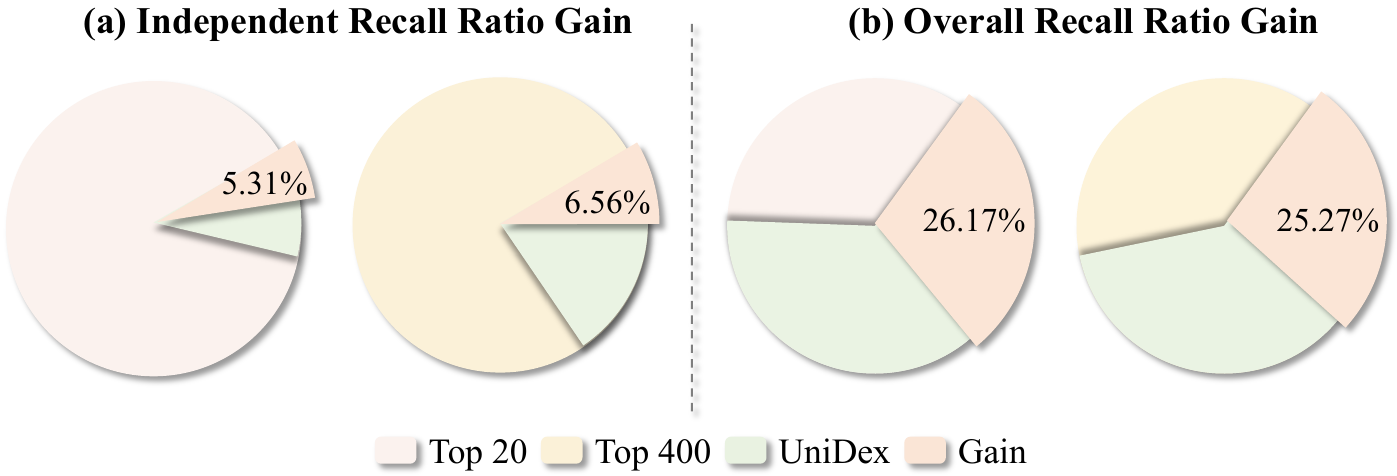}
\caption{The variations in recall ratio following the implementation of our comprehensive model, UniDex.
}
\label{pic:ratio}
\end{figure}

\begin{figure}[t]
\centering
\includegraphics[width=8.7cm]{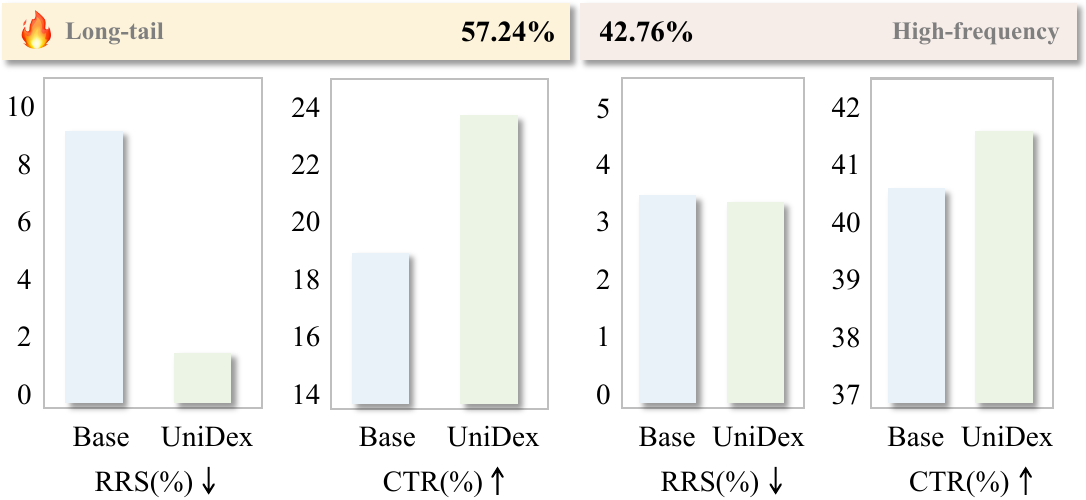}
\caption{Data analysis of Long-tail and High-frequency Queries based on improvements in Independent Recall Ratio for the Top 20. The top percentage reflects the contribution levels of queries at various frequencies. RRS denotes the Relevant Result Scarcity rate, while CTR refers to the Click-through Rate of the results.
}
\label{pic:fre_ana}
\end{figure}

\noindent \textbf{Enhancing Retrieval Effectiveness.}  The recall proportion during the fine-grained ranking phase reflects the retrieval effectiveness of the model. We examine two types of recall ratios: the overall recall ratio, which includes any video retrieved through inverted indexing, and the independent recall ratio, which counts a video only if it is retrieved exclusively through this method. As illustrated in Figure \ref{pic:ratio}, UniDex demonstrates enhancements in both recall ratios. Notably, there is an increase of over 25\% in the overall recall ratio. This significant improvement can be attributed to UniDex's ability to greatly enhance the generalization capability of inverted indexing, allowing it to retrieve more videos that overlap with semantic retrieval. In contrast, earlier term-based inverted indexing techniques were limited in capturing deeper search semantics. Furthermore, when we expand the number of ranked videos from the top 20 to the top 400, UniDex's independent recall rate increased by an additional 1.25\%. This indicates that integrating unified semantic modeling with inverted indexing provides a powerful complementary enhancement to the existing search pipeline, enabling the retrieval of more videos that were previously unretrievable.

\begin{figure}[t]
\centering
\includegraphics[width=8cm]{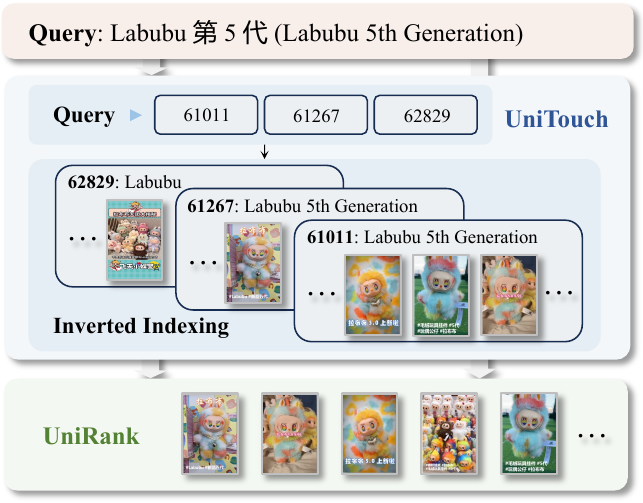}
\caption{The visualization of the UniDex retrieval process.
}
\label{pic:case}
\end{figure}

\noindent \textbf{Advancing Long-tail Semantic Characterization.} We conduct a detailed analysis of the sources of improvements in independent recall ratios, categorizing them by queries of varying frequencies.  As shown in Figure \ref{pic:fre_ana}, long-tail queries contribute 14.48\% (57.24\% vs 42.76\%) more than high-frequency queries. We also evaluate the Relevant Result Scarcity (RRS) rate and Click-through Rate (CTR) separately for both long-tail and high-frequency queries. A lower RRS score indicates better relevance of the retrieval results, while a higher CTR score suggests greater user satisfaction. Our findings reveal that UniDex significantly lowers the RRS and boosts the CTR for long-tail queries, highlighting its superior semantic representation capabilities compared to term-based inverted indexing. While UniDex demonstrates improvements over the baseline in representing high-frequency queries, these advancements are less pronounced compared to the significant gains observed for long-tail queries. This discrepancy may be due to the fact that term-based inverted indexing methods excel at capturing the semantics of high-frequency queries. However, long-tail queries demand a deeper textual understanding, which renders simple term-matching methods inadequate. By introducing unified semantic modeling, our UniDex model enhances the semantic representation capabilities of inverted indexing, resulting in notable improvements for long-tail queries. This enhanced semantic capability for long-tail queries addresses more nuanced search needs, thereby playing a crucial role for the platform.

\noindent \textbf{Balancing Generalization and Matching.}  To provide a clearer understanding of how UniDex operates, we perform a case analysis, with the results presented in Figure \ref{pic:case}. When the user inputs the query ``Labubu 5th Generation'', UniDex first activates the UniTouch module, converting the query into a series of SIDs ({\it e.g.}, ``61011, 61276, 62829''). These SIDs are then fed into a pre-stored inverted indexing that links SIDs to corresponding video lists, resulting in a set of candidate videos. Finally, the UniRank module is employed to rank the videos within this candidate set. In this case, most videos linked to SID-61011 are not retrieved by the traditional inverted indexing because the term ``Labubu'' does not appear in the video content. However, UniDex can utilize unified semantic modeling to understand the various linguistic expressions of the entity ``Labubu'', allowing it to retrieve videos with text descriptions that include the Chinese name for Labubu. The enhancement of semantic generalization raises the upper limit of inverted index retrieval, creating new opportunities for business innovation and iteration. Meanwhile, we also observe that SID-62829 employs a term omission strategy similar to term-based inverted indexing. Most of the videos associated with it are related to Labubu but lack crucial information such as ``5th''. This capability becomes particularly crucial when the candidate set contains only a few items related to the query, as it allows for the retrieval of as many vaguely relevant videos as possible. Consequently, UniDex retains the term-matching advantages of term-based inverted indexing while enhancing generalization.

\section{Conclusion}
The paper presents \textbf{UniDex}, a novel model-based approach that fundamentally transforms inverted indexing from a traditional term-based paradigm to a unified semantic modeling framework. By effectively replacing complex manual term-matching strategies with a streamlined semantic model, UniDex significantly reduces maintenance overhead and enhances retrieval capabilities through improved semantic generalization. Our successful deployment of UniDex in Kuaishou's short-video search systems, supported by extensive online A/B testing, demonstrates its practical effectiveness and scalability, serving millions of users efficiently. UniDex not only addresses the limitations of existing approaches but also opens new avenues for future innovations in search technology.

\bibliographystyle{ACM-Reference-Format}
\bibliography{main}


\begin{thebibliography}{48}


\ifx \showCODEN    \undefined \def \showCODEN     #1{\unskip}     \fi
\ifx \showDOI      \undefined \def \showDOI       #1{#1}\fi
\ifx \showISBNx    \undefined \def \showISBNx     #1{\unskip}     \fi
\ifx \showISBNxiii \undefined \def \showISBNxiii  #1{\unskip}     \fi
\ifx \showISSN     \undefined \def \showISSN      #1{\unskip}     \fi
\ifx \showLCCN     \undefined \def \showLCCN      #1{\unskip}     \fi
\ifx \shownote     \undefined \def \shownote      #1{#1}          \fi
\ifx \showarticletitle \undefined \def \showarticletitle #1{#1}   \fi
\ifx \showURL      \undefined \def \showURL       {\relax}        \fi
\providecommand\bibfield[2]{#2}
\providecommand\bibinfo[2]{#2}
\providecommand\natexlab[1]{#1}
\providecommand\showeprint[2][]{arXiv:#2}

\bibitem[Bai et~al\mbox{.}(2020)]%
        {sparterm2020}
\bibfield{author}{\bibinfo{person}{Yang Bai}, \bibinfo{person}{Xiaoguang Li}, \bibinfo{person}{Gang Wang}, \bibinfo{person}{Chaoliang Zhang}, \bibinfo{person}{Lifeng Shang}, \bibinfo{person}{Jun Xu}, \bibinfo{person}{Zhaowei Wang}, \bibinfo{person}{Fangshan Wang}, {and} \bibinfo{person}{Qun Liu}.} \bibinfo{year}{2020}\natexlab{}.
\newblock \showarticletitle{SparTerm: Learning Term-based Sparse Representation for Fast Text Retrieval}.
\newblock \bibinfo{journal}{\emph{arXiv preprint arXiv:2010.00768}} (\bibinfo{year}{2020}).
\newblock


\bibitem[Bengio et~al\mbox{.}(2013)]%
        {bengio2013estimating}
\bibfield{author}{\bibinfo{person}{Yoshua Bengio}, \bibinfo{person}{Nicholas L{\'e}onard}, {and} \bibinfo{person}{Aaron Courville}.} \bibinfo{year}{2013}\natexlab{}.
\newblock \showarticletitle{Estimating or Propagating Gradients Through Stochastic Neurons for Conditional Computation}.
\newblock \bibinfo{journal}{\emph{arXiv preprint arXiv:1308.3432}} (\bibinfo{year}{2013}).
\newblock


\bibitem[Bruch et~al\mbox{.}(2024)]%
        {bruch2024efficient}
\bibfield{author}{\bibinfo{person}{Sebastian Bruch}, \bibinfo{person}{Franco~Maria Nardini}, \bibinfo{person}{Cosimo Rulli}, {and} \bibinfo{person}{Rossano Venturini}.} \bibinfo{year}{2024}\natexlab{}.
\newblock \showarticletitle{Efficient Inverted Indexes for Approximate Retrieval over Learned Sparse Representations}. In \bibinfo{booktitle}{\emph{Proceedings of the 47th International ACM SIGIR Conference on Research and Development in Information Retrieval}}. \bibinfo{pages}{152--162}.
\newblock


\bibitem[Chen et~al\mbox{.}(2024)]%
        {chen2024generalizing}
\bibfield{author}{\bibinfo{person}{Haonan Chen}, \bibinfo{person}{Zhicheng Dou}, \bibinfo{person}{Kelong Mao}, \bibinfo{person}{Jiongnan Liu}, {and} \bibinfo{person}{Ziliang Zhao}.} \bibinfo{year}{2024}\natexlab{}.
\newblock \showarticletitle{Generalizing Conversational Dense Retrieval via {LLM}-Cognition Data Augmentation}. In \bibinfo{booktitle}{\emph{Proceedings of the 62nd Annual Meeting of the Association for Computational Linguistics (Volume 1: Long Papers)}}. \bibinfo{pages}{2700--2718}.
\newblock


\bibitem[Chen et~al\mbox{.}(2025)]%
        {chen2025unisearch}
\bibfield{author}{\bibinfo{person}{Jiahui Chen}, \bibinfo{person}{Xiaoze Jiang}, \bibinfo{person}{Zhibo Wang}, \bibinfo{person}{Quanzhi Zhu}, \bibinfo{person}{Junyao Zhao}, \bibinfo{person}{Feng Hu}, \bibinfo{person}{Kang Pan}, \bibinfo{person}{Ao Xie}, \bibinfo{person}{Maohua Pei}, \bibinfo{person}{Zhiheng Qin}, {et~al\mbox{.}}} \bibinfo{year}{2025}\natexlab{}.
\newblock \showarticletitle{UniSearch: Rethinking Search System with a Unified Generative Architecture}.
\newblock \bibinfo{journal}{\emph{arXiv preprint arXiv:2509.06887}} (\bibinfo{year}{2025}).
\newblock


\bibitem[Dai and Callan(2019)]%
        {dai2019deeper}
\bibfield{author}{\bibinfo{person}{Zhuyun Dai} {and} \bibinfo{person}{Jamie Callan}.} \bibinfo{year}{2019}\natexlab{}.
\newblock \showarticletitle{Deeper Text Understanding for IR with Contextual Neural Language Modeling}. In \bibinfo{booktitle}{\emph{Proceedings of the 42nd International ACM SIGIR Conference on Research and Development in Information Retrieval}}. \bibinfo{pages}{985--988}.
\newblock


\bibitem[Devlin et~al\mbox{.}(2019)]%
        {devlin-etal-2019-bert}
\bibfield{author}{\bibinfo{person}{Jacob Devlin}, \bibinfo{person}{Ming-Wei Chang}, \bibinfo{person}{Kenton Lee}, {and} \bibinfo{person}{Kristina Toutanova}.} \bibinfo{year}{2019}\natexlab{}.
\newblock \showarticletitle{{BERT}: Pre-training of Deep Bidirectional Transformers for Language Understanding}. In \bibinfo{booktitle}{\emph{Proceedings of the 2019 Conference of the North {A}merican Chapter of the Association for Computational Linguistics: Human Language Technologies, Volume 1 (Long and Short Papers)}}. \bibinfo{pages}{4171--4186}.
\newblock


\bibitem[Fkih(2023)]%
        {fkih2023enhancing}
\bibfield{author}{\bibinfo{person}{Fethi Fkih}.} \bibinfo{year}{2023}\natexlab{}.
\newblock \showarticletitle{Enhancing item-based collaborative filtering by users’ similarities injection and low-quality data handling}.
\newblock \bibinfo{journal}{\emph{Data \& Knowledge Engineering}}  \bibinfo{volume}{144} (\bibinfo{year}{2023}), \bibinfo{pages}{102126}.
\newblock


\bibitem[Formal et~al\mbox{.}(2021a)]%
        {formal2021splade}
\bibfield{author}{\bibinfo{person}{Thibault Formal}, \bibinfo{person}{Benjamin Piwowarski}, {and} \bibinfo{person}{St{\'e}phane Clinchant}.} \bibinfo{year}{2021}\natexlab{a}.
\newblock \showarticletitle{SPLADE: Sparse Lexical and Expansion Model for First Stage Ranking}. In \bibinfo{booktitle}{\emph{Proceedings of the 44th International ACM SIGIR Conference on Research and Development in Information Retrieval}}. \bibinfo{pages}{2288--2292}.
\newblock


\bibitem[Formal et~al\mbox{.}(2021b)]%
        {formal2021splade_v2}
\bibfield{author}{\bibinfo{person}{Thibault Formal}, \bibinfo{person}{Benjamin Piwowarski}, {and} \bibinfo{person}{St\'{e}phane Clinchant}.} \bibinfo{year}{2021}\natexlab{b}.
\newblock \showarticletitle{SPLADE: Sparse Lexical and Expansion Model for First Stage Ranking}. In \bibinfo{booktitle}{\emph{Proceedings of the 44th International ACM SIGIR Conference on Research and Development in Information Retrieval}}. \bibinfo{pages}{2288–2292}.
\newblock


\bibitem[Gao et~al\mbox{.}(2023)]%
        {gao2023alleviating}
\bibfield{author}{\bibinfo{person}{Chongming Gao}, \bibinfo{person}{Kexin Huang}, \bibinfo{person}{Jiawei Chen}, \bibinfo{person}{Yuan Zhang}, \bibinfo{person}{Biao Li}, \bibinfo{person}{Peng Jiang}, \bibinfo{person}{Shiqi Wang}, \bibinfo{person}{Zhong Zhang}, {and} \bibinfo{person}{Xiangnan He}.} \bibinfo{year}{2023}\natexlab{}.
\newblock \showarticletitle{Alleviating Matthew Effect of Offline Reinforcement Learning in Interactive Recommendation}. In \bibinfo{booktitle}{\emph{Proceedings of the 46th International ACM SIGIR Conference on Research and Development in Information Retrieval}}. \bibinfo{pages}{238--248}.
\newblock


\bibitem[Gao and Callan(2022)]%
        {gao2021unsupervised}
\bibfield{author}{\bibinfo{person}{Luyu Gao} {and} \bibinfo{person}{Jamie Callan}.} \bibinfo{year}{2022}\natexlab{}.
\newblock \showarticletitle{Unsupervised Corpus Aware Language Model Pre-training for Dense Passage Retrieval}. In \bibinfo{booktitle}{\emph{Proceedings of the 60th Annual Meeting of the Association for Computational Linguistics (Volume 1: Long Papers)}}. \bibinfo{pages}{2843--2853}.
\newblock


\bibitem[Gao et~al\mbox{.}(2021a)]%
        {gao2021coil}
\bibfield{author}{\bibinfo{person}{Luyu Gao}, \bibinfo{person}{Zhuyun Dai}, {and} \bibinfo{person}{Jamie Callan}.} \bibinfo{year}{2021}\natexlab{a}.
\newblock \showarticletitle{{COIL}: Revisit Exact Lexical Match in Information Retrieval with Contextualized Inverted List}. In \bibinfo{booktitle}{\emph{Proceedings of the 2021 Conference of the North American Chapter of the Association for Computational Linguistics: Human Language Technologies}}. \bibinfo{pages}{3030--3042}.
\newblock


\bibitem[Gao et~al\mbox{.}(2021b)]%
        {gao2021complement}
\bibfield{author}{\bibinfo{person}{Luyu Gao}, \bibinfo{person}{Zhuyun Dai}, \bibinfo{person}{Tongfei Chen}, \bibinfo{person}{Zhen Fan}, \bibinfo{person}{Benjamin Van~Durme}, {and} \bibinfo{person}{Jamie Callan}.} \bibinfo{year}{2021}\natexlab{b}.
\newblock \showarticletitle{Complement Lexical Retrieval Model with Semantic Residual Embeddings}. In \bibinfo{booktitle}{\emph{European Conference on Information Retrieval}}. Springer, \bibinfo{pages}{146--160}.
\newblock


\bibitem[He et~al\mbox{.}(2021)]%
        {he2021purify}
\bibfield{author}{\bibinfo{person}{Yue He}, \bibinfo{person}{Yancheng Dong}, \bibinfo{person}{Peng Cui}, \bibinfo{person}{Yuhang Jiao}, \bibinfo{person}{Xiaowei Wang}, \bibinfo{person}{Ji Liu}, {and} \bibinfo{person}{Philip~S Yu}.} \bibinfo{year}{2021}\natexlab{}.
\newblock \showarticletitle{Purify and Generate: Learning Faithful Item-to-Item Graph from Noisy User-Item Interaction Behaviors}. In \bibinfo{booktitle}{\emph{Proceedings of the 27th ACM SIGKDD Conference on Knowledge Discovery \& Data Mining}}. \bibinfo{pages}{3002--3010}.
\newblock


\bibitem[Jiang et~al\mbox{.}(2022)]%
        {jiang2022xlm}
\bibfield{author}{\bibinfo{person}{Xiaoze Jiang}, \bibinfo{person}{Yaobo Liang}, \bibinfo{person}{Weizhu Chen}, {and} \bibinfo{person}{Nan Duan}.} \bibinfo{year}{2022}\natexlab{}.
\newblock \showarticletitle{XLM-K: Improving Cross-Lingual Language Model Pre-training with Multilingual Knowledge}. In \bibinfo{booktitle}{\emph{Proceedings of the AAAI conference on artificial intelligence}}, Vol.~\bibinfo{volume}{36}. \bibinfo{pages}{10840--10848}.
\newblock


\bibitem[Karpukhin et~al\mbox{.}(2020)]%
        {karpukhin2020dense}
\bibfield{author}{\bibinfo{person}{Vladimir Karpukhin}, \bibinfo{person}{Barlas Oguz}, \bibinfo{person}{Sewon Min}, \bibinfo{person}{Patrick Lewis}, \bibinfo{person}{Ledell Wu}, \bibinfo{person}{Sergey Edunov}, \bibinfo{person}{Danqi Chen}, {and} \bibinfo{person}{Wen-tau Yih}.} \bibinfo{year}{2020}\natexlab{}.
\newblock \showarticletitle{Dense Passage Retrieval for Open-Domain Question Answering}. In \bibinfo{booktitle}{\emph{Proceedings of the 2020 Conference on Empirical Methods in Natural Language Processing (EMNLP)}}. \bibinfo{pages}{6769--6781}.
\newblock


\bibitem[Khattab and Zaharia(2020)]%
        {khattab2020colbert}
\bibfield{author}{\bibinfo{person}{Omar Khattab} {and} \bibinfo{person}{Matei Zaharia}.} \bibinfo{year}{2020}\natexlab{}.
\newblock \showarticletitle{ColBERT: Efficient and Effective Passage Search via Contextualized Late Interaction over BERT}. In \bibinfo{booktitle}{\emph{Proceedings of the 43rd International ACM SIGIR Conference on Research and Development in Information Retrieval}}. \bibinfo{pages}{39--48}.
\newblock


\bibitem[Lassance et~al\mbox{.}(2024)]%
        {lassance2024splade}
\bibfield{author}{\bibinfo{person}{Carlos Lassance}, \bibinfo{person}{Herv{\'e} D{\'e}jean}, \bibinfo{person}{Thibault Formal}, {and} \bibinfo{person}{St{\'e}phane Clinchant}.} \bibinfo{year}{2024}\natexlab{}.
\newblock \showarticletitle{SPLADE-v3: New baselines for SPLADE}.
\newblock \bibinfo{journal}{\emph{arXiv preprint arXiv:2403.06789}} (\bibinfo{year}{2024}).
\newblock


\bibitem[Lee et~al\mbox{.}(2021)]%
        {lee2021EWGS}
\bibfield{author}{\bibinfo{person}{Junghyup Lee}, \bibinfo{person}{Dohyung Kim}, {and} \bibinfo{person}{Bumsub Ham}.} \bibinfo{year}{2021}\natexlab{}.
\newblock \showarticletitle{Network Quantization with Element-wise Gradient Scaling}. In \bibinfo{booktitle}{\emph{2021 IEEE/CVF Conference on Computer Vision and Pattern Recognition (CVPR)}}, Vol.~\bibinfo{volume}{1}. \bibinfo{pages}{6444--6453}.
\newblock


\bibitem[Lee et~al\mbox{.}(2019)]%
        {lee2019latent}
\bibfield{author}{\bibinfo{person}{Kenton Lee}, \bibinfo{person}{Ming-Wei Chang}, {and} \bibinfo{person}{Kristina Toutanova}.} \bibinfo{year}{2019}\natexlab{}.
\newblock \showarticletitle{Latent Retrieval for Weakly Supervised Open Domain Question Answering}. In \bibinfo{booktitle}{\emph{Proceedings of the 57th Annual Meeting of the Association for Computational Linguistics}}. \bibinfo{pages}{6086--6096}.
\newblock


\bibitem[Lin and Ma(2021)]%
        {lin2021few}
\bibfield{author}{\bibinfo{person}{Jimmy Lin} {and} \bibinfo{person}{Xueguang Ma}.} \bibinfo{year}{2021}\natexlab{}.
\newblock \showarticletitle{A Few Brief Notes on DeepImpact, COIL, and a Conceptual Framework for Information Retrieval Techniques}.
\newblock \bibinfo{journal}{\emph{arXiv preprint arXiv:2106.14807}} (\bibinfo{year}{2021}).
\newblock


\bibitem[Liu et~al\mbox{.}(2024)]%
        {li2024llama2vec}
\bibfield{author}{\bibinfo{person}{Zheng Liu}, \bibinfo{person}{Chaofan Li}, \bibinfo{person}{Shitao Xiao}, \bibinfo{person}{Yingxia Shao}, {and} \bibinfo{person}{Defu Lian}.} \bibinfo{year}{2024}\natexlab{}.
\newblock \showarticletitle{{L}lama2{V}ec: Unsupervised Adaptation of Large Language Models for Dense Retrieval}. In \bibinfo{booktitle}{\emph{Proceedings of the 62nd Annual Meeting of the Association for Computational Linguistics (Volume 1: Long Papers)}}. \bibinfo{pages}{3490--3500}.
\newblock


\bibitem[Luan et~al\mbox{.}(2021)]%
        {luan2021sparse}
\bibfield{author}{\bibinfo{person}{Yi Luan}, \bibinfo{person}{Jacob Eisenstein}, \bibinfo{person}{Kristina Toutanova}, {and} \bibinfo{person}{Michael Collins}.} \bibinfo{year}{2021}\natexlab{}.
\newblock \showarticletitle{Sparse, Dense, and Attentional Representations for Text Retrieval Open Access}.
\newblock \bibinfo{journal}{\emph{Transactions of the Association for Computational Linguistics}}  \bibinfo{volume}{9} (\bibinfo{year}{2021}), \bibinfo{pages}{329--345}.
\newblock


\bibitem[MacAvaney et~al\mbox{.}(2020)]%
        {MacAvaney_2020}
\bibfield{author}{\bibinfo{person}{Sean MacAvaney}, \bibinfo{person}{Franco~Maria Nardini}, \bibinfo{person}{Raffaele Perego}, \bibinfo{person}{Nicola Tonellotto}, \bibinfo{person}{Nazli Goharian}, {and} \bibinfo{person}{Ophir Frieder}.} \bibinfo{year}{2020}\natexlab{}.
\newblock \showarticletitle{Expansion via Prediction of Importance with Contextualization}.
\newblock \bibinfo{journal}{\emph{Proceedings of the 43rd International ACM SIGIR Conference on Research and Development in Information Retrieval}} (\bibinfo{date}{Jul} \bibinfo{year}{2020}).
\newblock
\showISBNx{9781450380164}


\bibitem[Magnani et~al\mbox{.}(2022)]%
        {magnani2022semantic}
\bibfield{author}{\bibinfo{person}{Alessandro Magnani}, \bibinfo{person}{Feng Liu}, \bibinfo{person}{Suthee Chaidaroon}, \bibinfo{person}{Sachin Yadav}, \bibinfo{person}{Praveen Reddy~Suram}, \bibinfo{person}{Ajit Puthenputhussery}, \bibinfo{person}{Sijie Chen}, \bibinfo{person}{Min Xie}, \bibinfo{person}{Anirudh Kashi}, \bibinfo{person}{Tony Lee}, {et~al\mbox{.}}} \bibinfo{year}{2022}\natexlab{}.
\newblock \showarticletitle{Semantic Retrieval at Walmart}. In \bibinfo{booktitle}{\emph{Proceedings of the 28th ACM SIGKDD Conference on Knowledge Discovery and Data Mining}}. \bibinfo{pages}{3495--3503}.
\newblock


\bibitem[Mentzer et~al\mbox{.}(2024)]%
        {mentzer2023finite}
\bibfield{author}{\bibinfo{person}{Fabian Mentzer}, \bibinfo{person}{David Minnen}, \bibinfo{person}{Eirikur Agustsson}, {and} \bibinfo{person}{Michael Tschannen}.} \bibinfo{year}{2024}\natexlab{}.
\newblock \showarticletitle{Finite Scalar Quantization: VQ-VAE Made Simple}.
\newblock \bibinfo{journal}{\emph{The Twelfth International Conference on Learning Representations}}.
\newblock


\bibitem[Nogueira et~al\mbox{.}(2019a)]%
        {nogueira2019doc2query}
\bibfield{author}{\bibinfo{person}{Rodrigo Nogueira}, \bibinfo{person}{Jimmy Lin}, {and} \bibinfo{person}{AI Epistemic}.} \bibinfo{year}{2019}\natexlab{a}.
\newblock \showarticletitle{From doc2query to docTTTTTquery}.
\newblock \bibinfo{journal}{\emph{Online preprint}}  \bibinfo{volume}{6} (\bibinfo{year}{2019}).
\newblock


\bibitem[Nogueira et~al\mbox{.}(2019b)]%
        {nogueira2019document}
\bibfield{author}{\bibinfo{person}{Rodrigo Nogueira}, \bibinfo{person}{Wei Yang}, \bibinfo{person}{Jimmy Lin}, {and} \bibinfo{person}{Kyunghyun Cho}.} \bibinfo{year}{2019}\natexlab{b}.
\newblock \showarticletitle{Document Expansion by Query Prediction}.
\newblock \bibinfo{journal}{\emph{arXiv preprint arXiv:1904.08375}} (\bibinfo{year}{2019}).
\newblock


\bibitem[Oord et~al\mbox{.}(2018)]%
        {oord2018representation}
\bibfield{author}{\bibinfo{person}{Aaron van~den Oord}, \bibinfo{person}{Yazhe Li}, {and} \bibinfo{person}{Oriol Vinyals}.} \bibinfo{year}{2018}\natexlab{}.
\newblock \showarticletitle{Representation Learning with Contrastive Predictive Coding}.
\newblock \bibinfo{journal}{\emph{arXiv preprint arXiv:1807.03748}} (\bibinfo{year}{2018}).
\newblock


\bibitem[Pang et~al\mbox{.}(2017)]%
        {pang2017DeepRank}
\bibfield{author}{\bibinfo{person}{Liang Pang}, \bibinfo{person}{Yanyan Lan}, \bibinfo{person}{Jiafeng Guo}, \bibinfo{person}{Jun Xu}, \bibinfo{person}{Jingfang Xu}, {and} \bibinfo{person}{Xueqi Cheng}.} \bibinfo{year}{2017}\natexlab{}.
\newblock \showarticletitle{DeepRank: A New Deep Architecture for Relevance Ranking in Information Retrieval}. In \bibinfo{booktitle}{\emph{Proceedings of the 2017 ACM on Conference on Information and Knowledge Management}}. \bibinfo{pages}{257–266}.
\newblock


\bibitem[Qi et~al\mbox{.}(2025)]%
        {qi2025data}
\bibfield{author}{\bibinfo{person}{Xiaohua Qi}, \bibinfo{person}{Renda Li}, \bibinfo{person}{Long Peng}, \bibinfo{person}{Qiang Ling}, \bibinfo{person}{Jun Yu}, \bibinfo{person}{Ziyi Chen}, \bibinfo{person}{Peng Chang}, \bibinfo{person}{Mei Han}, {and} \bibinfo{person}{Jing Xiao}.} \bibinfo{year}{2025}\natexlab{}.
\newblock \showarticletitle{Data-free Knowledge Distillation with Diffusion Models}.
\newblock \bibinfo{journal}{\emph{arXiv preprint arXiv:2504.00870}} (\bibinfo{year}{2025}).
\newblock


\bibitem[Qu et~al\mbox{.}(2021)]%
        {qu2020rocketqa}
\bibfield{author}{\bibinfo{person}{Yingqi Qu}, \bibinfo{person}{Yuchen Ding}, \bibinfo{person}{Jing Liu}, \bibinfo{person}{Kai Liu}, \bibinfo{person}{Ruiyang Ren}, \bibinfo{person}{Wayne~Xin Zhao}, \bibinfo{person}{Daxiang Dong}, \bibinfo{person}{Hua Wu}, {and} \bibinfo{person}{Haifeng Wang}.} \bibinfo{year}{2021}\natexlab{}.
\newblock \showarticletitle{{R}ocket{QA}: An Optimized Training Approach to Dense Passage Retrieval for Open-Domain Question Answering}. In \bibinfo{booktitle}{\emph{Proceedings of the 2021 Conference of the North American Chapter of the Association for Computational Linguistics: Human Language Technologies}}. \bibinfo{pages}{5835--5847}.
\newblock


\bibitem[Ren et~al\mbox{.}(2021)]%
        {ren2021pair}
\bibfield{author}{\bibinfo{person}{Ruiyang Ren}, \bibinfo{person}{Shangwen Lv}, \bibinfo{person}{Yingqi Qu}, \bibinfo{person}{Jing Liu}, \bibinfo{person}{Wayne~Xin Zhao}, \bibinfo{person}{QiaoQiao She}, \bibinfo{person}{Hua Wu}, \bibinfo{person}{Haifeng Wang}, {and} \bibinfo{person}{Ji-Rong Wen}.} \bibinfo{year}{2021}\natexlab{}.
\newblock \showarticletitle{{PAIR}: Leveraging Passage-Centric Similarity Relation for Improving Dense Passage Retrieval}. In \bibinfo{booktitle}{\emph{Findings of the Association for Computational Linguistics: ACL-IJCNLP 2021}}. \bibinfo{pages}{2173--2183}.
\newblock


\bibitem[Roelleke and Wang(2008)]%
        {roelleke2008tf}
\bibfield{author}{\bibinfo{person}{Thomas Roelleke} {and} \bibinfo{person}{Jun Wang}.} \bibinfo{year}{2008}\natexlab{}.
\newblock \showarticletitle{TF-IDF uncovered: a study of theories and probabilities}. In \bibinfo{booktitle}{\emph{Proceedings of the 31st Annual International ACM SIGIR Conference on Research and Development in Information Retrieval}}. \bibinfo{pages}{435--442}.
\newblock


\bibitem[Santhanam et~al\mbox{.}(2022)]%
        {santhanam2021colbertv2}
\bibfield{author}{\bibinfo{person}{Keshav Santhanam}, \bibinfo{person}{Omar Khattab}, \bibinfo{person}{Jon Saad-Falcon}, \bibinfo{person}{Christopher Potts}, {and} \bibinfo{person}{Matei Zaharia}.} \bibinfo{year}{2022}\natexlab{}.
\newblock \showarticletitle{{C}ol{BERT}v2: Effective and Efficient Retrieval via Lightweight Late Interaction}. In \bibinfo{booktitle}{\emph{Proceedings of the 2022 Conference of the North American Chapter of the Association for Computational Linguistics: Human Language Technologies}}. \bibinfo{pages}{3715--3734}.
\newblock


\bibitem[Shi et~al\mbox{.}(2025)]%
        {shi2025retrieval}
\bibfield{author}{\bibinfo{person}{Teng Shi}, \bibinfo{person}{Jun Xu}, \bibinfo{person}{Xiao Zhang}, \bibinfo{person}{Xiaoxue Zang}, \bibinfo{person}{Kai Zheng}, \bibinfo{person}{Yang Song}, {and} \bibinfo{person}{Han Li}.} \bibinfo{year}{2025}\natexlab{}.
\newblock \showarticletitle{Retrieval Augmented Generation with Collaborative Filtering for Personalized Text Generation}. In \bibinfo{booktitle}{\emph{Proceedings of the 48th International ACM SIGIR Conference on Research and Development in Information Retrieval}}. \bibinfo{pages}{1294--1304}.
\newblock


\bibitem[Wang et~al\mbox{.}(2025)]%
        {wang2025personalized}
\bibfield{author}{\bibinfo{person}{Zhibo Wang}, \bibinfo{person}{Xiaoze Jiang}, \bibinfo{person}{Zhiheng Qin}, {and} \bibinfo{person}{Enyun Yu}.} \bibinfo{year}{2025}\natexlab{}.
\newblock \showarticletitle{Personalized Query Auto-Completion for Long and Short-Term Interests with Adaptive Detoxification Generation}. In \bibinfo{booktitle}{\emph{Proceedings of the 31st ACM SIGKDD Conference on Knowledge Discovery and Data Mining V. 2}}. \bibinfo{pages}{5018--5028}.
\newblock


\bibitem[Xiong et~al\mbox{.}(2021)]%
        {xiong2020approximate}
\bibfield{author}{\bibinfo{person}{Lee Xiong}, \bibinfo{person}{Chenyan Xiong}, \bibinfo{person}{Ye Li}, \bibinfo{person}{Kwok-Fung Tang}, \bibinfo{person}{Jialin Liu}, \bibinfo{person}{Paul Bennett}, \bibinfo{person}{Junaid Ahmed}, {and} \bibinfo{person}{Arnold Overwijk}.} \bibinfo{year}{2021}\natexlab{}.
\newblock \showarticletitle{Approximate Nearest Neighbor Negative Contrastive Learning for Dense Text Retrieval}. In \bibinfo{booktitle}{\emph{International Conference on Learning Representations}}.
\newblock


\bibitem[Yamada et~al\mbox{.}(2021)]%
        {yamada-etal-2021-efficient}
\bibfield{author}{\bibinfo{person}{Ikuya Yamada}, \bibinfo{person}{Akari Asai}, {and} \bibinfo{person}{Hannaneh Hajishirzi}.} \bibinfo{year}{2021}\natexlab{}.
\newblock \showarticletitle{Efficient Passage Retrieval with Hashing for Open-domain Question Answering}. In \bibinfo{booktitle}{\emph{Proceedings of the 59th Annual Meeting of the Association for Computational Linguistics and the 11th International Joint Conference on Natural Language Processing (Volume 2: Short Papers)}}. \bibinfo{pages}{979--986}.
\newblock


\bibitem[Yang et~al\mbox{.}(2017)]%
        {yang2017anserini}
\bibfield{author}{\bibinfo{person}{Peilin Yang}, \bibinfo{person}{Hui Fang}, {and} \bibinfo{person}{Jimmy Lin}.} \bibinfo{year}{2017}\natexlab{}.
\newblock \showarticletitle{Anserini: Enabling the Use of Lucene for Information Retrieval Research}. In \bibinfo{booktitle}{\emph{Proceedings of the 40th International ACM SIGIR Conference on Research and Development in Information Retrieval}}. \bibinfo{pages}{1253--1256}.
\newblock


\bibitem[Yang et~al\mbox{.}(2024)]%
        {yang2024trisampler}
\bibfield{author}{\bibinfo{person}{Zhen Yang}, \bibinfo{person}{Zhou Shao}, \bibinfo{person}{Yuxiao Dong}, {and} \bibinfo{person}{Jie Tang}.} \bibinfo{year}{2024}\natexlab{}.
\newblock \showarticletitle{TriSampler: A Better Negative Sampling Principle for Dense Retrieval}. In \bibinfo{booktitle}{\emph{Proceedings of the AAAI Conference on Artificial Intelligence}}, Vol.~\bibinfo{volume}{38}. \bibinfo{pages}{9269--9277}.
\newblock


\bibitem[Zhan et~al\mbox{.}(2021a)]%
        {zhan2021jointly}
\bibfield{author}{\bibinfo{person}{Jingtao Zhan}, \bibinfo{person}{Jiaxin Mao}, \bibinfo{person}{Yiqun Liu}, \bibinfo{person}{Jiafeng Guo}, \bibinfo{person}{Min Zhang}, {and} \bibinfo{person}{Shaoping Ma}.} \bibinfo{year}{2021}\natexlab{a}.
\newblock \showarticletitle{Jointly Optimizing Query Encoder and Product Quantization to Improve Retrieval Performance}. In \bibinfo{booktitle}{\emph{Proceedings of the 30th ACM International Conference on Information \& Knowledge Management}}. \bibinfo{pages}{2487--2496}.
\newblock


\bibitem[Zhan et~al\mbox{.}(2021b)]%
        {zhan2021optimizing}
\bibfield{author}{\bibinfo{person}{Jingtao Zhan}, \bibinfo{person}{Jiaxin Mao}, \bibinfo{person}{Yiqun Liu}, \bibinfo{person}{Jiafeng Guo}, \bibinfo{person}{Min Zhang}, {and} \bibinfo{person}{Shaoping Ma}.} \bibinfo{year}{2021}\natexlab{b}.
\newblock \showarticletitle{Optimizing Dense Retrieval Model Training with Hard Negatives}. In \bibinfo{booktitle}{\emph{Proceedings of the 44th International ACM SIGIR Conference on Research and Development in Information Retrieval}}. \bibinfo{pages}{1503--1512}.
\newblock


\bibitem[Zhan et~al\mbox{.}(2022)]%
        {zhan2022learning}
\bibfield{author}{\bibinfo{person}{Jingtao Zhan}, \bibinfo{person}{Jiaxin Mao}, \bibinfo{person}{Yiqun Liu}, \bibinfo{person}{Jiafeng Guo}, \bibinfo{person}{Min Zhang}, {and} \bibinfo{person}{Shaoping Ma}.} \bibinfo{year}{2022}\natexlab{}.
\newblock \showarticletitle{Learning Discrete Representations via Constrained Clustering for Effective and Efficient Dense Retrieval}. In \bibinfo{booktitle}{\emph{Proceedings of the Fifteenth ACM International Conference on Web Search and Data Mining}}. \bibinfo{pages}{1328--1336}.
\newblock


\bibitem[Zhang et~al\mbox{.}(2022)]%
        {zhang2022multi}
\bibfield{author}{\bibinfo{person}{Shunyu Zhang}, \bibinfo{person}{Yaobo Liang}, \bibinfo{person}{Ming Gong}, \bibinfo{person}{Daxin Jiang}, {and} \bibinfo{person}{Nan Duan}.} \bibinfo{year}{2022}\natexlab{}.
\newblock \showarticletitle{Multi-View Document Representation Learning for Open-Domain Dense Retrieval}. In \bibinfo{booktitle}{\emph{Proceedings of the 60th Annual Meeting of the Association for Computational Linguistics (Volume 1: Long Papers)}}. \bibinfo{pages}{5990--6000}.
\newblock


\bibitem[Zhao et~al\mbox{.}(2021)]%
        {zhao2020sparta}
\bibfield{author}{\bibinfo{person}{Tiancheng Zhao}, \bibinfo{person}{Xiaopeng Lu}, {and} \bibinfo{person}{Kyusong Lee}.} \bibinfo{year}{2021}\natexlab{}.
\newblock \showarticletitle{{SPARTA}: Efficient Open-Domain Question Answering via Sparse Transformer Matching Retrieval}. In \bibinfo{booktitle}{\emph{Proceedings of the 2021 Conference of the North American Chapter of the Association for Computational Linguistics: Human Language Technologies}}. \bibinfo{pages}{565--575}.
\newblock


\bibitem[Zhao et~al\mbox{.}(2022)]%
        {zhao2022improving}
\bibfield{author}{\bibinfo{person}{Xu Zhao}, \bibinfo{person}{Yi Ren}, \bibinfo{person}{Ying Du}, \bibinfo{person}{Shenzheng Zhang}, {and} \bibinfo{person}{Nian Wang}.} \bibinfo{year}{2022}\natexlab{}.
\newblock \showarticletitle{Improving Item Cold-start Recommendation via Model-agnostic Conditional Variational Autoencoder}. In \bibinfo{booktitle}{\emph{Proceedings of the 45th International ACM SIGIR Conference on Research and Development in Information Retrieval}}. \bibinfo{pages}{2595--2600}.
\newblock


\end{thebibliography}



 \end{document}